\documentclass[twocolumn,showpacs,preprintnumbers,amsmath,prl,amssymb,superscriptaddress,floatfix]{revtex4}

\usepackage{graphicx}
\usepackage{dcolumn}
\usepackage{bm}
\usepackage{mathrsfs}
\usepackage{multirow}

\def\braxion{( \Gamma _a / \Gamma _{\gamma} )}
\def\breq{\frac{\Gamma _a}{\Gamma _{\gamma}}}
\def\cpd{\rm{day^{-1}kg^{-1}keV^{-1}}}
\def\nuebar{\rm{\bar{\nu_e}}}
\def\nue{\rm{\nu_e}}

\def\gagg{g_{a \gamma \gamma}}
\def\gaee{g_{a e e }}
\def\gs0{g^0_{a N N }}
\def\gv1{g^1_{a N N }}
\def\dk2g{\Gamma_{\gamma \gamma}}
\def\dkee{\Gamma_{e e}}
\def\pn2dg{\rm{pn \to d \gamma}}

\begin{document}

\preprint{AS-TEXONO/06-05}

\title{
Search of axions 
from a nuclear power reactor
with a high-purity germanium detector
}


\newcommand{\as}{Institute of Physics, Academia Sinica, Taipei 115, Taiwan.}
\newcommand{\ntu}{Department of Physics, National Taiwan University,
Taipei 106, Taiwan.}
\newcommand{\ihep}{Institute of High Energy Physics,
Chinese Academy of Science, Beijing 100039, China.}
\newcommand{\thu}{Department of Engineering Physics, Tsing Hua University,
Beijing 100084, China.}
\newcommand{\ciae}{Department of Nuclear Physics,
Institute of Atomic Energy, Beijing 102413, China.}
\newcommand{\metu}{Department of Physics,
Middle East Technical University, Ankara 06531, Turkey.}
\newcommand{\corr}{htwong@phys.sinica.edu.tw}

\author{ H.M.~Chang }  \affiliation{ \as } \affiliation{ \ntu }
\author{ H.T.~Wong } \altaffiliation[Corresponding Author: ]{ \corr } \affiliation{ \as }
\author{ M.H.~Chou } \affiliation{ \as }
\author{ M.~Deniz } \affiliation{ \as } \affiliation{ \metu }
\author{ H.X.~Huang } \affiliation{ \as } \affiliation{ \ciae }
\author{ F.S.~Lee } \affiliation{ \as }
\author{ H.B.~Li }  \affiliation{ \as }
\author{ J.~Li }  \affiliation{ \ihep } \affiliation{ \thu }
\author{ H.Y~Liao } \affiliation{ \as } \affiliation{ \ntu }
\author{ S.T.~Lin } \affiliation{ \as } \affiliation{ \ntu }
\author{ V.~Singh } \affiliation{ \as } 
\author{ S.C.~Wu } \affiliation{ \as }
\author{ B.~Xin } \affiliation{ \ciae }

\collaboration{TEXONO Collaboration}

\noaffiliation


\date{\today}

\begin{abstract}

A search of axions produced in 
nuclear transitions was performed at the 
Kuo-Sheng Nuclear Power Station
with a high-purity germanium detector
of mass 1.06~kg at a distance of 28~m from 
the 2.9~GW reactor core.
The expected experimental signatures were mono-energetic
lines produced by their
Primakoff or Compton conversions 
at the detector.  
Based on 459.0/96.3 days of Reactor ON/OFF data,
no evidence of axion emissions were observed and
constraints on the 
couplings $\gagg$ and $\gaee$ versus axion mass $m _a$
within the framework of invisible axion models
were placed.  
The KSVZ and DFSZ models can be excluded for 
$\rm{10^4 ~ eV \alt m _a \alt 10^6 ~eV}$.
Model-independent constraints on
$\gagg \cdot \gv1 < 7.7 \times 10^{-9} ~ {\rm GeV^{-2}}$ 
for $m_{a} \alt 10^5 ~ {\rm eV}$ 
and 
$\gaee \cdot \gv1 < 1.3 \times 10^{-10} $ 
for $m_{a} \alt 10^6 ~ {\rm eV}$ 
at 90\% confidence level 
were derived. 
This experimental approach provides a unique probe 
for axion mass at the keV$-$MeV range 
not accessible to the other techniques.

\end{abstract}

\pacs{
14.80.Mz, 29.40.-n, 28.41.-i
}

\maketitle

\section{I. Introduction}

The axions ($a$)~\cite{pdg}  were
proposed in the 1970's as solution to
the strong CP problem $-$
the near-perfect cancellations
between the QCD vacuum angle and the quark mass matrix,
as constrained by experimental bounds on
the neutron electric dipole moments.
The interactions of the 
axions with matter (photons, electrons
and nucleons) can be described 
by the effective Lagrangian~\cite{pdg,avignone}
\begin{eqnarray}
\label{eq::lag}
  \mathscr{L}_{int} & = & 
 \gagg \phi_a \vec{E} \cdot \vec{B}  \nonumber \\
&& + ~ i \gaee \phi_a 
 \overline{\phi}_e \gamma_{5} \phi_e  \nonumber \\
&& + ~ i \phi _a \overline{\psi}_N \gamma_{5} 
( \gs0 + \gv1 \tau_3 ) \psi_N  ~ ,
\end{eqnarray}
where $\phi _a$, ($\vec{E}, \vec{B}$),
$\psi _e$ and 
$\psi _N  = \left( \begin{array}{c} p \\ n \end{array} \right) $ 
represent
respectively the axion, electromagnetic,
electron and nucleon fields.
The couplings of the axions
to the photons and electrons are parametrized by
$\gagg$ and $\gaee$, while
$\gs0$ and $\gv1$ are their
isoscalar and isovector couplings to the nucleons.

A generic feature of the axion models is that
all the coupling constants as well as the
axion mass ($m _a$) are inversely
proportional to the symmetry breaking scale ($f _a$).
The original ``Peccei-Quinn-Weinberg-Wilczek'' (PQWW)
model~\cite{pqww} took $f _a$ to be the electroweak scale, 
implying $m _a$ of the order of $\sim$100~keV.
This has been tested and
excluded after extensive efforts.
Current research programs~\cite{exptreview}
focus on larger $f _a$ in the 
``invisible axion models'',
the two popular variants
of which are the DFSZ (or GUT)~\cite{dfsz}
and KSVZ (or hadronic) models~\cite{ksvz}.
The axion couplings with matter
within the framework of these models were evaluated 
and discussed in details in 
Refs.~\cite{kaplan}\&\cite{srednicki}.

The light-mass axion is a well-motivated 
dark matter candidate. 
Cosmological and astrophysical 
arguments~\cite{pdg,raffelt} constrain the axion mass
to be $\rm{10^{-6} ~ eV < m_{a} < 10^{-2} ~ eV}$, 
but the bounds are model-dependent and 
with large uncertainties. 
Experiments have been performed
to look for dark matter axions 
as well as those produced in the sun,
power reactors and radioactive 
nuclear transitions.

All previous reactor experiments~\cite{raxgg,raxee} 
focused on the searches of the
PQWW axions via their decays, 
and contributed much to 
exclude their existence. 
In this article, we present 
results on a new search over
a broad axion mass range,
using an alternative detection strategy
through its interactions with matter.
This detection scheme was successfully
used in a previous experiment 
using radioactive isotope
as axion source~\cite{avignone}.

\section{II. Reactor as Axion Source}

The axions are pseudoscalar particles and have
quantum numbers like those of magnetic photons. 
It can be emitted through magnetic transitions 
in radioactive gamma-decays~\cite{donnelly}.
Nuclear power reactors are 
powerful radioactive sources and are
therefore potential axion sources as well.
Axions can be emitted in competition with
the photons as a result of neutron captures
\begin{equation}
n+(Z,A) \to (Z,A+1)+\gamma / a 
\end{equation}
or nuclear de-excitations
\begin{equation}
  (Z,A)^{*} \to (Z,A)+\gamma / a ~~ .
\end{equation}

There are six prominent channels of magnetic
gamma-transitions at typical 
nuclear reactors, as listed in Table~\ref{tabrate}. 
Thermal neutron captures 
on the $^{10}$B in the control rods
and on proton in the cooling water produce 
$\alpha$+$^{7}$Li$^{*}$ and d+$\gamma$, respectively. 
Their photon fluxes ($\Phi _{\gamma}$),
in units of fission$^{-1}$ and GCi,
were evaluated by full
neutron transport simulations~\cite{rnue}.
The other sources of 
$^{91}$Y$^{*}$, $^{97}$Nb$^{*}$, 
$^{135}$Xe$^{*}$, and $^{137}$Ba$^{*}$ are 
all fission daughters. Their corresponding
$\Phi _{\gamma}$ were derived from standard
tables on fission yields~\cite{fyield} and cross-checked
by previous calculations~\cite{raxgg}.
For comparisons, the $\nuebar$~\cite{texonomagmom} 
and $\nue$-yields~\cite{rnue}
at reactors are about 7.2 and 
$\rm{\sim 10^{-3}~ fission^{-1}}$, respectively.

\begin{table}
\caption
{\label{tabrate}
A summary of magnetic transitions and their estimated
fluxes at a typical 2.9~GW power reactor.}
\begin{ruledtabular}
\begin{tabular}{lclcc}
Channel & $\rm{E _{\gamma}}$  
& Transitions & \multicolumn{2}{c}{$\Phi_{\gamma}$}\\
& (keV) & & (fission$^{-1}$) & (GCi) \\ 
\hline
p(n,$\gamma$)d & 2230 & Isovector M1 & 0.25 & 0.61 \\ 
$^{10}$B(n,$\alpha$)$^{7}$Li$^*$ & 478 & 
M1 $(\frac{1}{2}^{-}) \to (\frac{3}{2}^{-})$ & 0.28 & 0.68 \\ \hline
$^{91}$Y$^*$ & 555 & 
M4 $(\frac{9}{2}^{+}) \to (\frac{1}{2}^{-})$ & 0.024 & 0.058 \\
$^{97}$Nb$^*$ & 743 & 
M4 $(\frac{1}{2}^{-}) \to (\frac{9}{2}^{+})$ & 0.055 & 0.13 \\
$^{135}$Xe$^*$ & 526 & 
M4 $(\frac{11}{2}^{-}) \to (\frac{3}{2}^{+})$ & 0.0097 & 0.023 \\
$^{137}$Ba$^*$ & 662 & 
M4 $(\frac{11}{2}^{-}) \to (\frac{3}{2}^{+})$ & 0.0042 & 0.010 \\
\end{tabular}
\end{ruledtabular}
\end{table}

The axion flux ($\phi_{a}$) at 
a distance $L$ from a reactor core of
fission rate $R _f$ can be described by
\begin{equation}
\label{eq::aflux}
  \phi_{a} (L) ~ = ~ \frac{R _f \cdot \Phi _{\gamma}}{4 \pi L^2} \cdot
\breq \cdot P _{dk} \cdot P _{int} ~~ ,  
\end{equation}
where 
$\braxion$ is the branching ratio of axion emissions
in the transitions. 
It depends on the axion-nucleon couplings 
and the nuclear structures of the transitions. 
The probabilities of the axions surviving
the flight from reactor core to detector without decays
or interactions
are given, respectively, by
\begin{eqnarray}
 P _{dk}  & = & {\rm exp} [ - \frac{L \cdot m_{a}}{p _{a} \cdot \tau _{a}} ] \\
& {\rm and} & \nonumber \\
 P _{int} & = & {\rm exp} [ - L  \cdot \rho _{L} \cdot  \sigma _{int} ] ~~~ ,
\end{eqnarray}
where $m _{a}$, $\tau _{a}$, $p_{a}$, $E _{a}$ are
the axion mass, lifetime, momentum and total energy,
$\sigma _{int}$ is 
the axion interaction cross section with
matter at effective target
number density $\rho _{L}$. 

In particular, the axions can decay in flight via 
the emissions of 2$\gamma$
($\dk2g : a \rightarrow \gamma \gamma$) 
or $\rm{e^+ e^-}$ pairs
($\dkee : a \rightarrow e^+ e^-$).
Their decay rates are related to the
$\gagg$ and $\gaee$ couplings
by~\cite{donnelly}:
\begin{eqnarray}
\frac{1}{\dk2g} & = & \frac{64\pi}{g_{a\gamma\gamma}^{2}m_{a}^{3}} \\
& {\rm and} & \nonumber \\
\frac{1}{\dkee} & = & \frac{8\pi}{g_{aee}^{2}\sqrt{m_{a}^{2}-4m_{e}^{2}}} ~~~ .
\end{eqnarray}
The axion lifetime is then given by
\begin{equation}
\tau _a ~ = ~ \frac{1}{ \dk2g + \dkee }  ~~ .
\end{equation}


\section{III. Axion Detection and Experimental Setup}

\begin{figure}[hbt]
\includegraphics[width=8cm]{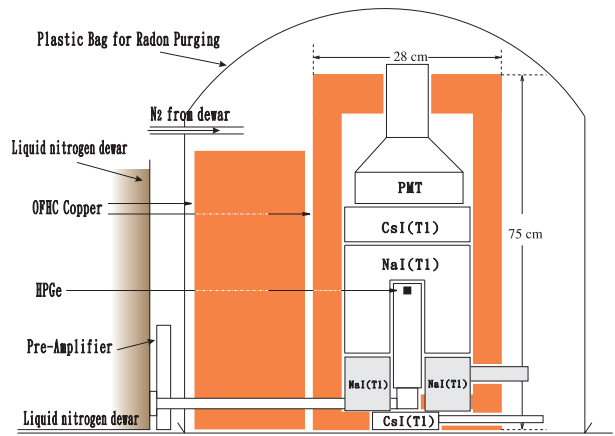}
\caption{
Schematic layout of the HPGe
with its anti-Compton detectors
as well as inner shieldings and
radon purge system.
}
\label{kshpge}
\end{figure}

Data were taken with
a high-purity germanium detector (HPGe) of
mass 1.06~kg
at the Kuo-Sheng (KS) Reactor Laboratory. 
The HPGe target and the associated anti-Compton (AC)
detectors as well as passive shieldings are depicted
in Figure~\ref{kshpge}.
The principal AC detector
was a well-shaped NaI(Tl) crystal scintillator
of mass 19.7~kg.
Descriptions of the experimental hardware and analysis procedures
were presented in details in 
Refs.~\cite{texonomagmom}\&\cite{texonodaq}.
The primary scientific goal was the search of
neutrino magnetic moments.
A physics threshold of 12~keV 
and a background level of 1~$\cpd$ comparable to 
those of underground dark matter experiments
were achieved.
The source-detector distance of KS was $L$=28~m
while $\rho _L$ was modeled to be
27.75~m of water and 0.25~m of lead.

\begin{figure}[hbt]
\includegraphics[width=8.0cm]{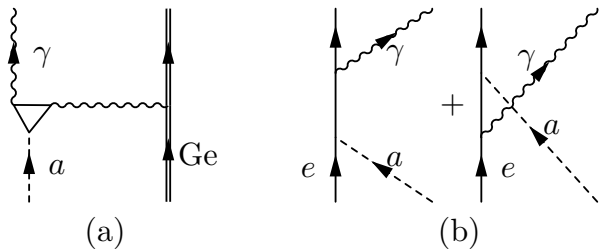}
\caption{\label{feyn}
Schematic diagrams of interactions of axions with
matter, via
(a) Primakoff and (b) Compton conversions.
}
\end{figure}

\begin{figure}[hbt]
\includegraphics[width=8cm]{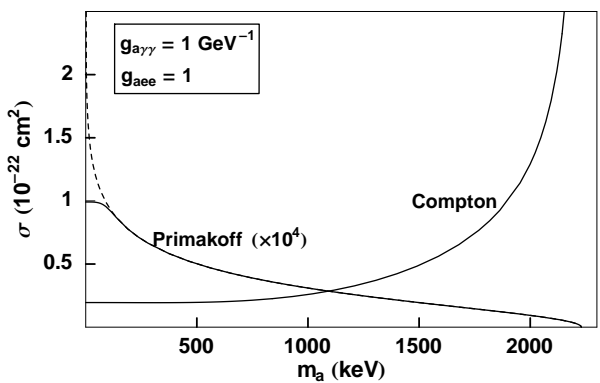}
\caption{
The variations of the Primakoff and
Compton conversion cross-sections with the
axion mass at $E _a = 2.23 ~ {\rm MeV}$,
using the normalizations of 
$\gagg = 1 ~ {\rm GeV ^{-1}}$
and $\gaee = 1$.
The dashed line represents the case where
the atomic screening effects are switched
off ($\chi = 1$).
}
\label{csvsma}
\end{figure}

The search strategies for reactor axions
with these unique HPGe data were inspired by
a previous experiment~\cite{avignone}
where a 15~kCi $\gamma$-source of $^{65}$Zn
was used as a potential axion source instead.
Two interaction
mechanisms of axions with matter
were studied: Primakoff and Compton conversions as 
shown schematically in Figures~\ref{feyn}a\&b.
These processes are respectively and independently
sensitive to $\gagg$ and $\gaee$.
Their cross-sections
were both given in Ref.~\cite{avignone} $-$
(a) Primakoff conversion on the nuclei: 
\begin{equation}
\label{eq::prim}
\sigma_{P} = \gagg ^{2} 
\frac{Z^{2} \alpha_{em}}{2}
\frac{1}{\beta} \left[ \frac{1+\beta^{2}}{2\beta} 
\ln \left[ \frac{1+\beta}{1-\beta}
\right] -1 \right] \chi  ~~  ,
\end{equation}
where $\alpha_{em}$ is the electromagnetic coupling,
$Z$ is the atomic number of the target, 
$\beta = p_{a} / E_{a}$ 
and $\chi$ is the atomic-screening correction factor
given in Eq.~20 of Ref.~\cite{avignone};
(b) Compton conversion on the electrons: 
\begin{eqnarray}
\label{eq::compt}
\sigma_{C} & = & \gaee^{2}
\frac{\pi \alpha_{em}}{8 \pi m_{e}^{2} p_{a}}
[ ~ \frac{2m_{e}^{2}(m_{e}+E_{a})y}{(m_{e}^{2}+y)^{2}} \nonumber \\
& & + ~
\frac{4m_{e}(m_{a}^{4}+2m_{a}^{2}m_{e}^{2}-4m_{e}^{2}E_{a}^{2})}
{y(m_{e}^{2}+y)} 
\nonumber \\
& & + ~
\frac{4m_{e}^{2}p_{a}^{2}+m_{a}^{4}}{p_{a}y} \ln
\frac{m_{e}+E_{a}+p_{a}}{m_{e}+E_{a}-p_{a}} ~ ] ~~ ,
\end{eqnarray}
where $y=2m_{e} E_{a}+m_{a}^{2}$.
The $m_a$ dependence of the two cross-sections
at the $\pn2dg$ transition energy of 
$E _a = 2.23 ~ {\rm MeV}$, 
using the normalizations of 
$\gagg = 1 ~ {\rm GeV ^{-1}}$
and $\gaee = 1$,
are illustrated in Figure~\ref{csvsma}. 

The total energy of the axions
can be fully converted 
by either of the two processes
into measurable ionization energy
in the HPGe, such that the experimental signatures
are the presence of mono-energetic lines 
at the known $E _a$'s during 
the Reactor ON periods.
In comparison, previous reactor-based axion experiments
studied instead the axion decay channels 
$\dk2g$~\cite{raxgg} or 
$\dkee$~\cite{raxee}.
They were therefore not sensitive to
the invisible axion regime where $m _a$
are very small and decays 
are kinematically blocked or suppressed.

\section{IV. Data Analysis}

The signal rates ($S _{P/C}$) for axion Primakoff/Compton 
Conversions in one kilogram of target mass
are given by
\begin{eqnarray}
\label{eq::rprim}
 S _{P}  & =  & \sigma _{P} \cdot \phi_{a} \cdot N \cdot \epsilon_{P} \\
& {\rm and} & \nonumber \\
 S_{C}  & =  & \sigma _{C} \cdot \phi_{a} \cdot
Z  \cdot N \cdot  \epsilon_{C} ~~ ,
\label{eq::rcmpt}
\end{eqnarray}
where $\epsilon_{P/C}$ are
the efficiencies
of full energy deposition at the HPGe detector,
$N$ is the number of atoms in the kilogram target,  
and $Z$ accounts for the
electron target number in Compton process.
The various efficiency factors were evaluated by full
simulations and listed in Table~\ref{tabresults}.
Full energy depositions for Compton conversion
at the HPGe detectors were due to
interactions only in Ge, such that only
$N$(Ge), $Z$(Ge)=32 
and $\epsilon _C (Ge)$ were involved
in the derivation of $S _C$.
However, photons from 
Primakoff conversions in both Ge and NaI could
contribute to $S _P$ 
as full-energy peaks at the HPGe,
such that there are terms
involving, respectively, [$N$(Ge), $\epsilon _P$(Ge)] 
and [$N$(NaI), $\epsilon _P$(NaI)].


\begin{figure}[hbt]
{\bf (a)}\\
\includegraphics[width=8cm]{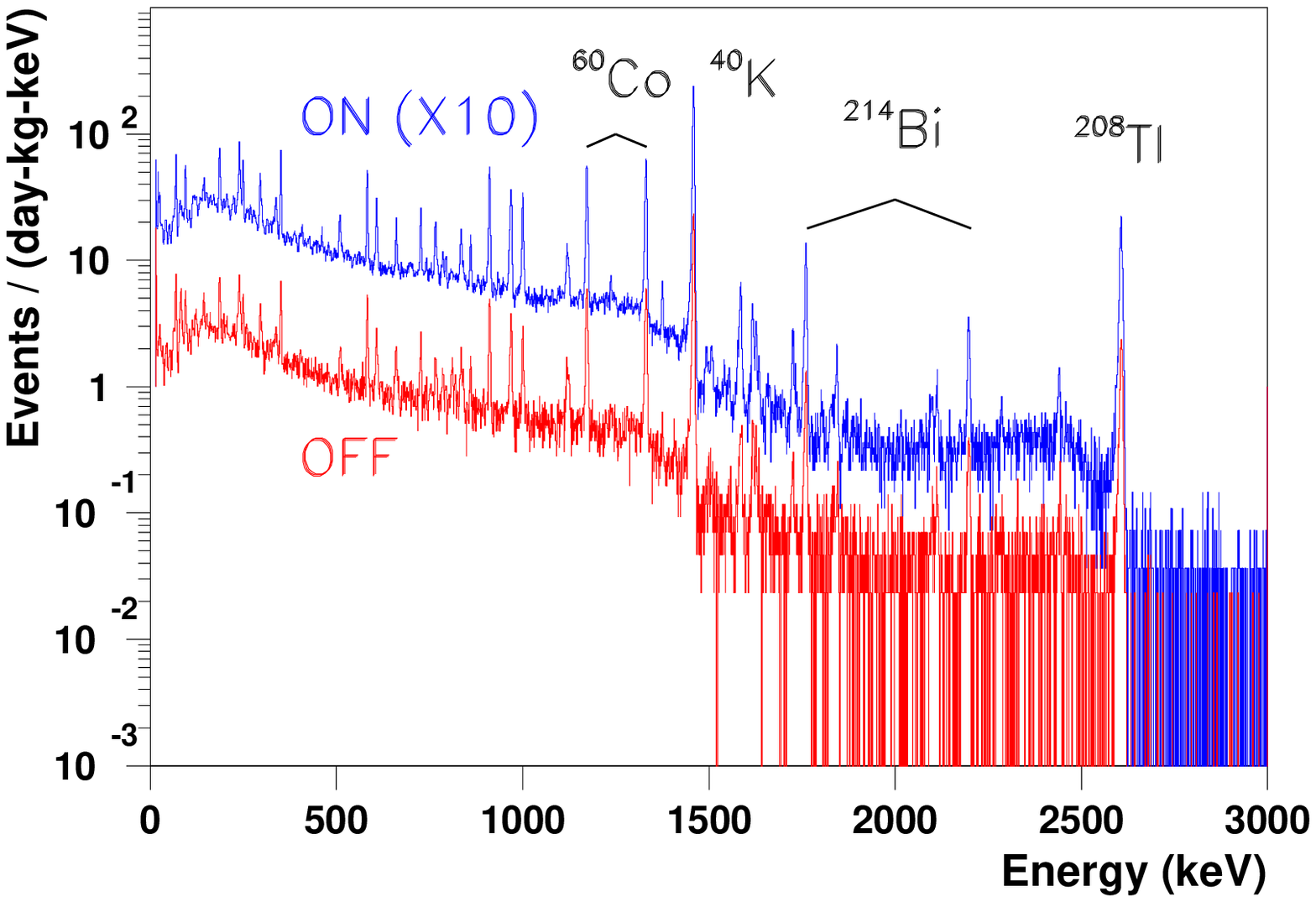}\\
{\bf (b)}\\
\includegraphics[width=8cm]{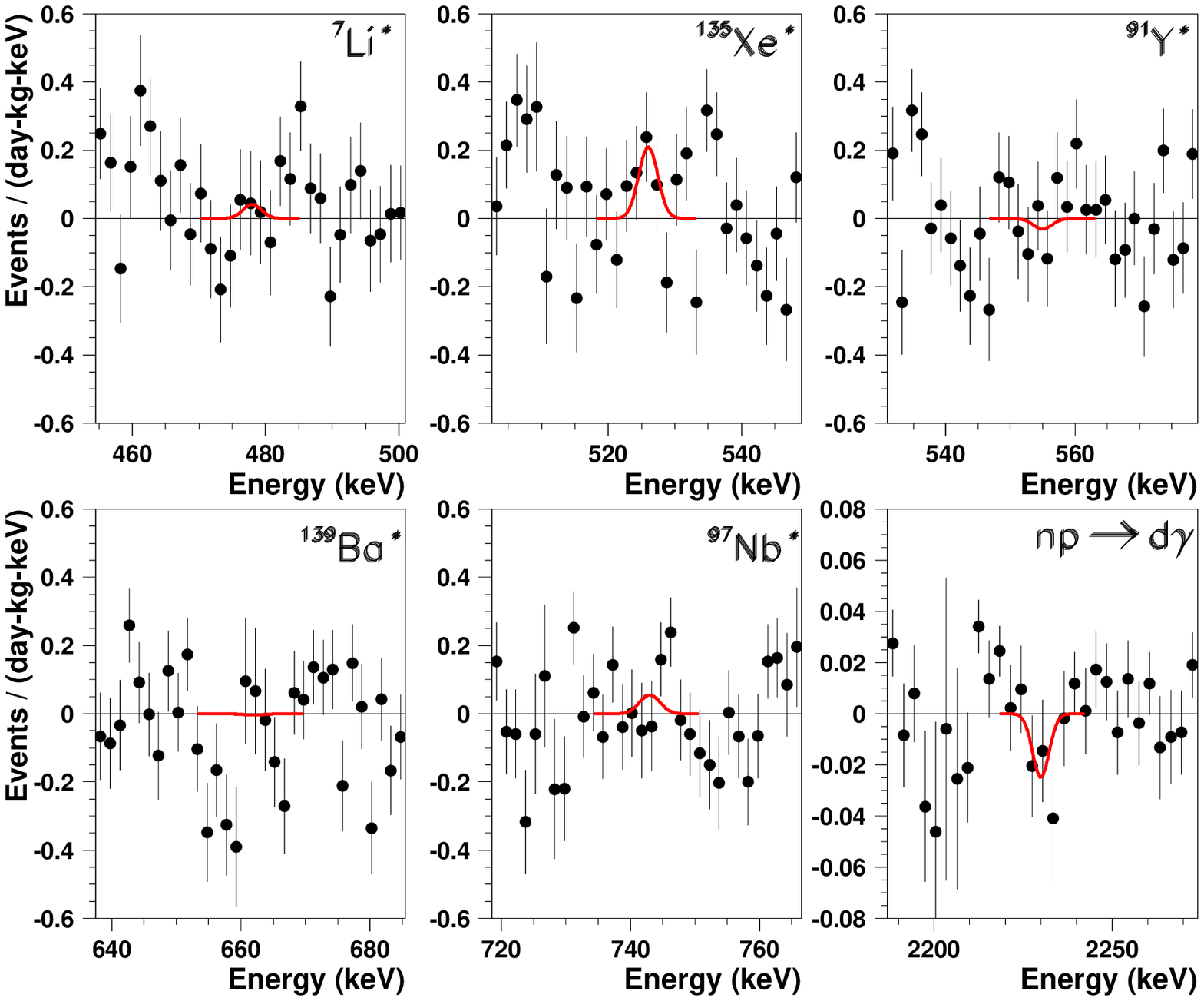}
\caption{
(a) The after-cut ON and OFF spectra
and (b) the residual spectra
for the six specific channels for
Period-III, with the best-fit Gaussian peaks 
overlaid.
}
\label{p3spect}
\end{figure}

\begin{table*}
\caption{
\label{tabresults}
Summary of the detector efficiencies for
Primakoff and Compton conversions on
axions of energy $E _{a}$, the measured
Reactor ON$-$OFF event rates at the signal regions and
the 90\% CL upper limits ($\rm{S_u}$).
}
\begin{ruledtabular}
\begin{tabular}{lr|ccc|ccc|c}
Channel & $E _{a}$
& $\epsilon _P (Ge)$ & $\epsilon _P (NaI)$ & $\epsilon _C (Ge)$ ~~ &
Period I & Period III & Combined & $S _u$ \\
& (keV)  & & & &
(day$^{-1}$kg$^{-1}$) & (day$^{-1}$kg$^{-1}$) & (day$^{-1}$kg$^{-1}$)
& (day$^{-1}$kg$^{-1}$) \\ \hline
$^7$Li$^*$ & 478    & 0.36&0.0048 & 0.61  &
-0.88$\pm$0.75 & 0.14$\pm$0.41 & -0.09$\pm$0.36 & $<$ 0.49 \\
$^{135}$Xe$^*$ & 526    & 0.34&0.0047 & 0.58  &
0.26$\pm$0.67 & 0.38$\pm$0.16 & 0.37$\pm$0.15 & $<$ 0.62 \\
$^{91}$Y$^*$ & 555    & 0.33&0.0044 & 0.58 &
-0.47$\pm$0.67 & -0.33$\pm$0.15 & -0.34$\pm$0.15 & $<$ 0.05 \\
$^{137}$Ba$^*$ & 662  & 0.30&0.0040 & 0.54 &
-0.46$\pm$0.62 & -0.02$\pm$0.50 & -0.19$\pm$0.39 & $<$ 0.46 \\
$^{97}$Nb$^*$ & 743    & 0.28&0.0037 & 0.53 &
0.14$\pm$0.55 & 0.22$\pm$0.37 & 0.19$\pm$0.31 & $<$ 0.69 \\
$\pn2dg$ & 2230   & 0.16&0.0020 & 0.37 &
-0.10$\pm$0.17 & -0.03$\pm$0.03 & -0.04$\pm$0.03 & $<$ 0.02 \\
\end{tabular}
\end{ruledtabular}
\end{table*}


Evidence of reactor axions would manifest as peaks 
at the known energies of Table~\ref{tabrate} in the
Reactor ON$-$OFF residual spectra in HPGe.
Following the naming conventions of Ref.~\cite{texonomagmom},
Periods-I (July 2001$-$April 2002)
and 
-III (Sept. 2004$-$Oct. 2005) 
with 180.1/52.7 
and 278.9/43.6 
days of 
the KS-HPGe Reactor ON/OFF data,
respectively, were used for analysis.
Candidate events were those uncorrelated with the anti-Compton
and Cosmic-Ray vetos and having pulse shapes consistent with
$\gamma$-events. Selections of these events and their
efficiencies were discussed in Ref.~\cite{texonomagmom}.
As illustrations, the Period-III 
ON/OFF background spectra 
and the ON$-$OFF residual spectra for the
six candidate lines
are depicted in Figures~\ref{p3spect}a\&b, respectively.
The background $\gamma$-lines were 
identified~\cite{texonomagmom} and indicated that
ambient radioactivity dominated.

The count rates and their errors 
of the various transitions
shown in Table~\ref{tabresults}
were derived by
best-fits of the residual spectra
to Gaussian lines at fixed $E_a$'s and
resolutions.
No excess were observed in all channels and 
upper limits of the signal rates ($S _u$) at
90\% confidence level(CL) were derived.
The most sensitive channel 
in terms of $S _u$ is 
the 2.23~MeV transition 
(RMS resolution 2.3~keV) 
in the np$\to$d$a$ interaction,
because of 
the lower background level 
compared to those 
at~$\sim$500~keV~(by~$\sim 10^{-2}$~\cite{texonomagmom}).

For completeness, 
searches were also performed 
at:~(a)~half $E_a$ to look for 
single-$\gamma$ absorption
in $\dk2g$ decays, and
(b)~full $E_a$ of the
individual ON/OFF spectra to look
for axion emissions from other
possible steady-state sources,
such as those from the sun~\cite{solarli7,moriyama}. 
No signals were observed
in both cases.

\section{V. Derivations of Axion Parameters}

The experimentally measured
upper limits $S _u$ of Table~\ref{tabresults}
can be translated to bounds among the
axion parameters: $m_a$, $\gagg$, $\gaee$,
$\gs0$ and $\gv1$.
Two approaches were adopted: (a) specific
models were used and tested, and 
(b) model-independent constraints among
the parameters were derived. 

\subsection{A. Branching Ratios for Axion Emissions}

The limits $S _u$ 
are related to reactor axion emission 
and detection via
\begin{equation}
\label{eq::slimit}
S _{P} ~ + ~ S _{C} ~ < ~ S _u ~~ .
\end{equation}
Both $S_P$ and $S_C$ depend on the reactor axion
flux, and thus the
branching ratios $\braxion$.

Starting from the interaction Lagrangian of Eq.~\ref{eq::lag},
the axion emission
branching ratio of the $\pn2dg$ isovector M1
transition can be
expressed as~\cite{donnelly,barroso}:
\begin{eqnarray}
\label{eq::pn2dg}
 ( \breq ) _{pn}  & \equiv  & \breq ( \pn2dg ) \nonumber \\
& = & ( \frac{1}{2 \pi \alpha_{em}} )
~ ( \frac{p_a}{p_{\gamma}} ) ^3 
~ ( \frac{\gv1}{\mu_1}  )^2  ~~ ,
\end{eqnarray}
while those for M$L$  
transitions in general are~\cite{avignone,donnelly}
\begin{eqnarray}
\label{eq::br}
\breq & = & ( \frac{1}{\pi \alpha_{em}} )
~ ( \frac{1}{1 + \delta ^2}  )
~ ( \frac{L}{L+1} )
~ ( \frac{p_a}{p_{\gamma}} ) ^{2L+1}  \nonumber \\
& & [ \frac{\gs0 \beta + \gv1}{(\mu_0 - \frac{1}{2} ) \beta 
+ ( \mu_1 - \eta )}]^2 ~~ .
\end{eqnarray}
In the formulae, $L$ is the multi-polarity of the 
transition,
$\delta$ is the known E($L+1$)/M$L$ mixing ratio, 
$\mu _0$ and $\mu _1$ 
are respectively the isoscalar and isovector 
magnetic moments which can be derived from the proton and neutron
magnetic moments ($\mu_p$ and $\mu_n$) via:
\begin{eqnarray}
\mu _0 & =  \mu _p + \mu _n  \simeq & 0.88  \\
& {\rm and} & \nonumber \\
\mu _1 & =  \mu _p - \mu _n  \simeq & 4.71  ~ .
\end{eqnarray}
The nuclear physics of the transitions are parametrized by
the matrix elements $\beta$ and $\eta$ defined as 
\begin{eqnarray}
 \beta & = & 
 \frac{ < J_f \mid\mid  {\displaystyle \sum_{i=1}^{A}}  
\sigma (i) \mid\mid J_i > }
{< J_f \mid\mid  {\displaystyle \sum_{i=1}^{A}} 
\sigma (i) \tau_3 (i) \mid\mid J_i >}  \\
& {\rm and} & \nonumber   \\
\eta & = & 
 - \frac{< J_f \mid\mid  {\displaystyle \sum_{i=1}^{A}} 
l(i) \tau_3 (i) \mid\mid J_i >}
{< J_f \mid\mid  {\displaystyle \sum_{i=1}^{A}} 
\sigma (i) \tau_3 (i) \mid\mid J_i >}  ~~ ,  
\end{eqnarray}
where $J_i$ and $J_f$ are the initial and final nuclear angular momentum 
in the transitions, while $l (i)$ and $\sigma (i)$ are the 
orbital angular momentum and nuclear spin operators.

\subsection{B. Invisible Axion Models}

The evaluations of $\braxion$ 
given in Eq.~\ref{eq::br} involve
modeling of the couplings $\gs0$ and $\gv1$.
Within the framework of
the invisible axion models,
these couplings are inversely proportional to
the symmetry breaking scale $f_a$ which
in turn is related to
$m_a$~\cite{kaplan,srednicki} via
\begin{equation}
\label{eq::ma}
m_a / {\rm eV} =
( \frac{1.3 \times 10^{7}}{ f_a  / {\rm GeV}} )
[ \frac{\sqrt{z}}{1+z} ]  ~~,
\end{equation}
such that the couplings depend linearly on $m_a$.

The formulae given in
Table~1 of Ref.~\cite{kaplan}
were adopted for the parametrizations of
these axion-nucleon couplings:
\begin{eqnarray}
\label{eq::kaplan}
\gs0 & = & C \cdot [ ~
\frac{( 3F-D )}{6} ~ ( X_u  - X_d - N_f ) \nonumber \\
& &  ~ + ~ \frac{S}{3} ~ ( X_u + 2 X_d - N_f )  ~ ] 
\end{eqnarray}
and
\begin{equation}
\gv1  =   C \cdot
\frac{( D+F )}{2} ~ [ ~ X_u - X_d - N_f ~ ( \frac{1-z}{1+z} ) ~ ] ~ ,
\end{equation}
where the factor common to both couplings is
\begin{equation}
C ~ = ~ 5.2 \times 10^{-8} ~ ( \frac{3}{N_f} ) [ m_a / {\rm eV} ]  ~~ .
\end{equation}
The terms $D \simeq 0.77$ and $F \simeq 0.48$~\cite{moriyama} 
are the reduced matrix elements for the octet
axial vector currents,
$S = 0.33 \pm 0.04$~\cite{svalue} denotes 
the flavor singlet axial charge, 
$N_f = 3$ is the number of families, 
$z = ( m_u / m_d ) \simeq 0.56$
is the ratio of the up-to-down
quark mass, 
while $X _u$ and $X _d$ represent
respectively the PQ charge of the u and d quarks.
The KSVZ and DFSZ models differ essentially
in their choices of $( X_u , X_d )$.

Previous evaluations of axion fluxes in
the stellar~\cite{haxton}
and solar~\cite{solarli7,moriyama} environment
adopted the KSVZ model where $X_u = X_d = 0$.
This model also specifies $\gaee = 0$ at
tree level such that
the results are only applicable to probe 
the axion-photon $\gagg$-couplings.
We extended the analysis to include also
the DFSZ model, which
allows finite $\gagg$- and $\gaee$-couplings.
The parameters $X_u$ and $X_d$ are positive-definite 
constrained by $X_u + X_d =1$.
The values of $X_u = X_d = 0.5$ were chosen 
for this analysis.
Defining
\begin{equation}
\label{eq::ann}
g^{0/1}_{aNN} 
 ~ \equiv ~  A^{0/1} ~ [ m_a / {\rm eV} ] ~~ ,
\end{equation}
the calculable numerical factors $A^{0/1}$ 
under both KSVZ and DFSZ models 
are tabulated in Table~\ref{axmodels}.
The isovector couplings $\gv1$ are the
same for both models with this specific 
choice of $( X_u , X_d )$.

\begin{table}
\caption
{\label{axmodels}
The calculated numerical factors
$A ^{0/1}$ of Eq.~\ref{eq::ann}
under the KSVZ and DFSZ invisible axion
models using the formulae of 
Ref.~\cite{kaplan}.
}
\begin{ruledtabular}
\begin{tabular}{lccc}
Model & $I$ & $A^I$ & Validity \\ \hline
KSVZ & 0 & $-3.5 \times 10^{-8}$
& \multirow{2}*{ \{ $ \begin{array}{l}  {\rm finite} ~ \gagg \\
\gaee = 0   \end{array}   $ \} }  \\
~~ $X_u$=$X_d$=0
& 1 &  $-2.8 \times 10^{-8}$ & \\ \hline
DFSZ & 0 & $-2.6 \times 10^{-8}$
& \multirow{2}*{ \{ $ \begin{array}{l}  {\rm finite } ~ \gagg  \\
\& ~ \gaee    \end{array}   $ \} }  \\
~~ $X_u$=$X_d$=0.5
& 1 & $-2.8 \times 10^{-8}$ & 
\end{tabular}
\end{ruledtabular}
\end{table}

\begin{table*}
\caption
{\label{nuclear}
A summary of the nuclear physics input in
the evaluations of $\braxion$. 
Calculated values of $( \beta , \eta )$ 
on the other isotopes from previous works
are included for comparisons.
The branching ratios $\braxion$ at $m_a$=1~eV
under the KSVZ and DFSZ invisible axion
models and the
quality factors (QF) relative to that for
the $\pn2dg$ channel are also shown. 
}
\begin{ruledtabular}
\begin{tabular}{lcc|cccl||cc|cc}
Channel & Unpaired & Transition & $\delta$ & $\beta$ & $\eta$ & 
Remarks & $\braxion ^\dagger$ & QF/(QF)$_{pn}$ 
&  $\braxion ^\dagger$ & QF/(QF)$_{pn}$ \\ 
& p/n & & & & & & \multicolumn{2}{c|}{KSVZ Model}  
& \multicolumn{2}{c}{DFSZ Model} \\ \hline
$\pn2dg$ & $-$ & M1  
& $-$ & $-$ & $-$ & Refs.~\cite{donnelly,raxgg,raxee} & 
$7.4 \times 10^{-16}$ & 1.00 &  
$7.4 \times 10^{-16}$ & 1.00  \\ 
$^{7}$Li$^*$ & p & M1 
& 0 & 1 & 0.5 & Ref.~\cite{solarli7} &
$4.0 \times 10^{-15}$ & 0.50 & 
$3.0 \times 10^{-15}$ & 0.43 \\
$^{91}$Y$^*$ & p & M4 
& 0 & 1 & -3 & 
\multirow{4}*{ \{  \(
\begin{array}{l} 
${\rm Inferred}$ \\ ${\rm  ~~from}$ \\ ${\rm Refs.}~\cite{avignone,haxton}$
\end{array} 
\) } 
&
$2.1 \times 10^{-15}$ & 0.33 & 
$1.5 \times 10^{-15}$ & 0.28 \\ 
$^{97}$Nb$^*$ & p & M4 
& 0 & 1 & -3 & &
$2.1 \times 10^{-15}$ & 0.13 &  
$1.5 \times 10^{-15}$ & 0.11 \\ 
$^{135}$Xe$^*$ & n & M4 
& 0 & -1 & 1 &  &
$1.6 \times 10^{-16}$ & 0.02 &  
$7.1 \times 10^{-18}$ & 0.003 \\ 
$^{137}$Ba$^*$ & n & M4 
& 0 & -1 & 1 & &
$1.6 \times 10^{-16}$ & 0.01 &  
$7.1 \times 10^{-18}$ & 0.003 \\ \hline
$^{65}$Cu$^*$ & p & M1 
&  0.44 & 1.81 & -6.59 & Ref.~\cite{avignone} &  
$1.0 \times 10^{-15}$ &  $-$ & $7.1 \times 10^{-16}$ & $-$ \\ 
$^{57}$Fe$^*$ & n & M1 
& 0.002 & -1.19 & 0.80 & Ref.~\cite{haxton} &  
$3.4 \times 10^{-16}$ &  $-$ & $2.2 \times 10^{-17}$ & $-$ \\ 
$^{55}$Mn$^*$ & p & M1 
& 0.052  & 0.79 & -3.74 & Ref.~\cite{haxton} &  
$8.5 \times 10^{-16}$ &  $-$ & $6.6 \times 10^{-16}$ & $-$ \\ 
$^{23}$Na$^*$ & p & M1 
& 0.058 & 0.88 & -1.20 & Ref.~\cite{haxton} &  
$1.9 \times 10^{-15}$ &  $-$ & $1.4 \times 10^{-15}$ & $-$ 
\end{tabular}
$^{\dagger}$ Evaluated at $m_a$=1~eV \hfill
\end{ruledtabular}
\end{table*}

Once $\gs0$ and $\gv1$
are fixed by the invisible axion models,
the evaluations of $\braxion$ depend on 
the nuclear physics
inputs: $\delta$, $\beta$ and $\eta$.
The values adopted for these parameters
are summarized in Table~\ref{nuclear}.
Among the transitions, the $\pn2dg$
is a pure isovector M1 process. 
Its branching ratio $\braxion _{pn}$, 
as given by Eq.~\ref{eq::pn2dg},
is independent of 
$( \delta , \beta , \eta )$ 
and was used in previous reactor axion
experiments~\cite{raxgg,raxee,donnelly}.
The $\rm{^7 Li ^*}$ transition is also
pre-dominantly M1.
The neutron shell is closed and the transition
is driven by the odd-proton, such that
$( \delta , \beta , \eta ) \simeq (0,1,0.5)$
was adopted as in Ref.~\cite{solarli7}.
There were no  calculations on $(\beta , \eta )$ for
the remaining
four fission daughter isotopes.
To make estimations, we project from the
results on other heavy isotopes~\cite{avignone,haxton},
also summarized in Table~\ref{nuclear}.
The matrix elements
$(\beta , \eta )$ for
heavy isotopes with unpaired proton and
neutron are taken to be
$\simeq (1,-3)$ and $\simeq (-1,1)$,
respectively.

\begin{figure}[hbt]
\includegraphics[width=8cm]{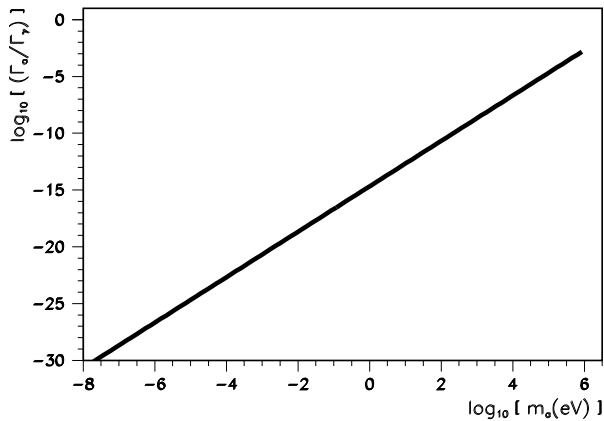}
\caption{
The variations of $\braxion$
with $m_a$ in both
KSVZ and DFSZ invisible axion models
as parametrized in Ref.~\cite{kaplan}.
The $\rm{^{135} Xe ^*}$ and $\rm{^{137} Ba ^*}$
channels have large uncertainties and were
discarded.  The width represents
the variations within the four remaining channels
and among the two models.
}
\label{brvsma}
\end{figure}

Once these $(\beta , \eta )$ assignments
are made, 
the branching ratios $\braxion$ 
can be readily evaluated 
with Eqs.~\ref{eq::pn2dg}$-$\ref{eq::ann}.
In particular, $\beta \leq  0$ for
the odd-neutron nuclei $\rm{^{135} Xe ^*}$
and $\rm{^{137} Ba ^*}$,
such that the $\gs0$ and $\gv1$ terms have 
opposite signs
and $\braxion$ can be vanishingly small within
the large nuclear physics uncertainties.
These two channels were discarded in subsequent
analysis.
The variations of $\braxion$ with $m_a$
of the four remaining channels
are depicted in Figure~\ref{brvsma}.
The differences within the channels
and between the KSVZ and DFSZ models
are small relative to the scale in the log-log plot,
and are represented by the width of the line.

The experimental sensitivities 
in $\gagg$ and $\gaee$ can be described by
a  quality factor (QF)
which is related to the $\gamma$-flux $\Phi _{\gamma}$
of Table~\ref{tabrate}
and the upper limit signal rates $S_u$ of
Table~\ref{tabresults} via:
\begin{equation}
\label{eq::qf}
\frac{1}{ g_{a \gamma \gamma / a e e} }  \propto 
{\rm QF} ~ \equiv ~
\sqrt{ \frac{ \Phi _{\gamma} \cdot \braxion }{S_u} } ~~ .
\end{equation}
Both $\braxion$ and the relative QF (with
respect to that of the $\pn2dg$ channel)
evaluated at
$m_a$=1~eV were shown 
in Table~\ref{nuclear} to illustrate
the relative strength of the various
channels.  
It can be seen that the
leading contribution to the sensitivities
is from the $\pn2dg$ channel. 

\subsection{C. Model-Dependent Limits}

The $\braxion$ estimates in  Table~\ref{nuclear} for 
$\pn2dg$ and $\rm{^{7} Li ^*}$
are accurate while those for the four heavy
fission isotopes are expected to have large uncertainties.
In addition, the relative QF values 
of Table~\ref{nuclear} indicate that the
$\pn2dg$ channel mostly
defines the sensitivities. 
Accordingly, we take the conservative approach
that the experimental limits on $\gagg$ and $\gaee$
were derived only from this channel.
The two detection channels were treated independently, 
via the relations 
\begin{equation}
\label{eq::s2limit}
S {_P} <  S _u  ~~~~~ \&   ~~~~~ S _{C} < S _u  ~~,
\end{equation}
assuming, respectively, 
$\gaee = 0$ and $\gagg = 0$. 
These produce less stringent bounds 
compared to those from the
convoluted case of Eq.~\ref{eq::slimit},
such that the results have general validity.

\begin{figure}[hbt]
\includegraphics[width=8cm]{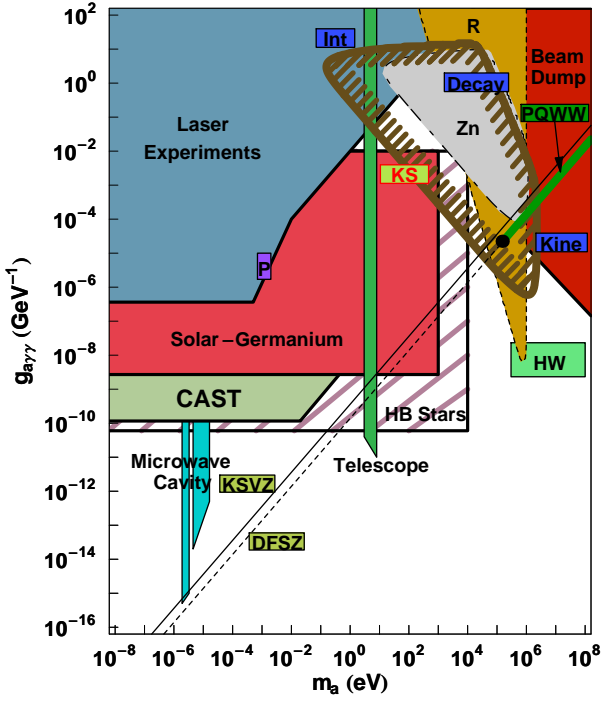}
\caption{
\label{ex-gagg}
Exclusion plots of $\gagg$ versus
$m_{a}$ for $\gaee$=0.
The limits from the KS experiment 
at 90\% CL are denoted by ``KS''.
They are derived by fixing $\braxion _{pn}$
with the KSVZ and DFSZ invisible axion models.
Predicted regions of the PQWW, KSVZ, DFSZ 
and HW models on the $( m_a , \gagg )$ plane
are overlaid.
The boundaries defined by  
``Int'', ``Decay'' and ``Kine''
are constraints
due to axions interactions with matter, decays in flights
and kinematics, respectively.
The bounds marked ``Zn''
are results from Ref.~\cite{avignone}, while
region labeled ``R'' are from
previous reactor axion experiments studying
$\dk2g$~\cite{raxgg}.
Results from the other axion experiments
using different 
techniques~\cite{beamdump,solarge,cast,cavity,laser} 
are displayed as colored blocks.
The astrophysical bounds~\cite{raffelt} 
are denoted by the striped region.
}
\end{figure}

\begin{figure}[hbt]
\includegraphics[width=8cm]{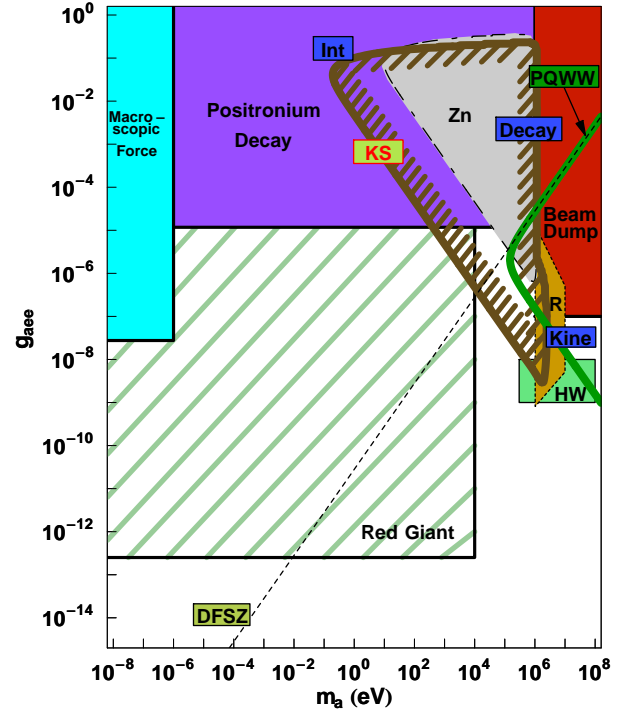}
\caption{
\label{ex-gaee}
Exclusion plots of $\gaee$ versus $m_{a}$
for $\gagg$=0.
Similar conventions as Figure~\ref{ex-gagg} are
adopted.
The ``KS'' bounds are derived by 
fixing $\braxion _{pn}$ with 
the DFSZ model.
Regions marked ``R'' was excluded by
experiment studying $\dkee$
at reactors~\cite{raxee}.
Bounds from
othe experiments~\cite{beamdump,asai,force}
are shown as colored blocks.
}
\end{figure}

The exclusion plot of
$\gagg$ versus $m_a$ is depicted
in Figure~\ref{ex-gagg}.
The model-dependence was introduced by fixing
$\braxion _{pn}$ as a function of $m _a$ 
via Eqs.~\ref{eq::pn2dg}\&\ref{eq::ann}.
The values of $\braxion _{pn}$
depend only on $\gv1$ 
and are therefore 
the same for both KSVZ and DFSZ
models, as shown in Table~\ref{nuclear}. 
Consequently, both models 
produce a common exclusion region, 
denoted by ``KS''.
On the other hand,
the KSVZ model specifies $\gaee = 0$
at tree level,
such that only the DFSZ model
can be meaningfully applied 
to define the exclusion region,
also labeled as ``KS'',
in the $\gaee$ versus $m_a$ plot
of Figure~\ref{ex-gaee}.

In both
Figures~\ref{ex-gagg}\&\ref{ex-gaee},
the vertical bounds labeled ``Kine'' at 2.23~MeV 
are due to the kinematical constraints
from the maximum $E _a$.
The sensitivities are suppressed at ``Decay''
for the large $ ( m_a , \gagg / \gaee )$ regions,
due to $\dk2g$ and $\dkee$ 
decays in flight. 
The lack of sensitivities at ``Int''
for large $\gagg \agt 20$~GeV$^{-1}$  and
$\gaee \agt 0.2$
are due to axion interactions in the matter
between reactor core and detector.
Limits marked ``R'' and ``Zn'' are 
respectively from previous reactor experiments
studying $\dk2g$~\cite{raxgg} or $\dkee$~\cite{raxee} and the 
radioactive source experiment~\cite{avignone}.
They were derived using the same modeling 
schemes on $\braxion$ as ``KS''.
The bounds from the KS reactor axion searches
improve on those of Ref.~\cite{avignone} by two orders
of magnitude, 
owing to enhanced axion flux, lower background
and larger data sample.


The KS results
define the global exclusion boundaries
in $\gagg$ for 
$10^3 ~{\rm eV} \alt m _a \alt 10^6 ~ {\rm eV}$
and in $\gaee$ for
$10^2 ~{\rm eV} \alt m _a \alt 10^6 ~ {\rm eV}$.
Astrophysics arguments on stellar cooling
and red giant yields~\cite{pdg,raffelt}
provide more stringent bounds
for $m_{a} \alt 10^4 ~{\rm eV}$ but
these are model-dependent.
They are represented by the striped regions
in Figures~\ref{ex-gagg}\&\ref{ex-gaee}, respectively.
Comparisons of the KS excluded regions
with the KSVZ/DFSZ predictions
on the $ ( m_a , \gagg / \gaee )$ planes
would rule out these models
at $10^4 ~{\rm eV} \alt m _a \alt 10^6 ~ {\rm eV}$.
An example of other existing model
predicting axion mass at the MeV range is
the HW model~\cite{hall}, also depicted
in the figures.

The experimental approach presented
in this article can probe the 
keV$-$MeV axion mass range 
which is not accessible to the other techniques.
At large $m_{a} \agt 10^6 ~{\rm eV}$,
the sensitivities in both 
$\gagg$  and $\gaee$ are dominated by
the accelerator-based ``beam dump'' 
experiments~\cite{beamdump}.
Exclusion boundaries at small $m_a$
are defined by: 
(a) for $\gagg$ $-$
the germanium~\cite{solarge} and CAST~\cite{cast} 
experiments studying solar axions, and the
axion dark matter searches with
microwave cavity~\cite{cavity};
and (b) for $\gaee$ $-$
the positronium decay~\cite{asai} 
and macroscopic force~\cite{force} experiments.
At $\gagg > 10^{-2} ~ {\rm GeV^{-1}}$, the
solar axion experiments are limited by
axion interactions inside the Sun~\cite{raffelt}. 
Part of this large-$\gagg$ region 
has been rejected by
the optical laser experiments~\cite{laser},
while the KS results contribute to probe 
and exclude a remaining hole
at $m_a \sim 10^{1} ~ {\rm eV}$.
For completeness, we mention also
the recent PVLAS experiment which reported
a finite light polarization rotation 
in vacuum with a transverse magnetic field~\cite{pvlas}.
This result was interpreted as 
the region ``P'' of finite ($m_a$,$\gagg$)
in Figure~\ref{ex-gagg},
well contradicted by many other experiments.
This would imply that it may not be appropriate
to analyze the PVLAS results using the
existing axion models.

\subsection{D. Model-Independent Constraints}

The KS exclusion regions of 
Figures~\ref{ex-gagg}\&\ref{ex-gaee}
were evaluated within 
the framework of the invisible
axion models.    
Alternatively, it is also instructive
to derive model-independent constraints 
among the axion parameters.
Following the reasonings of Section~V.C,
only the results from
the $\pn2dg$ channel were adopted,
and the two couplings $\gagg$ and $\gaee$ were
treated independently. 

\begin{figure}[hbt]
{\bf (a)}\\
\includegraphics[width=8cm]{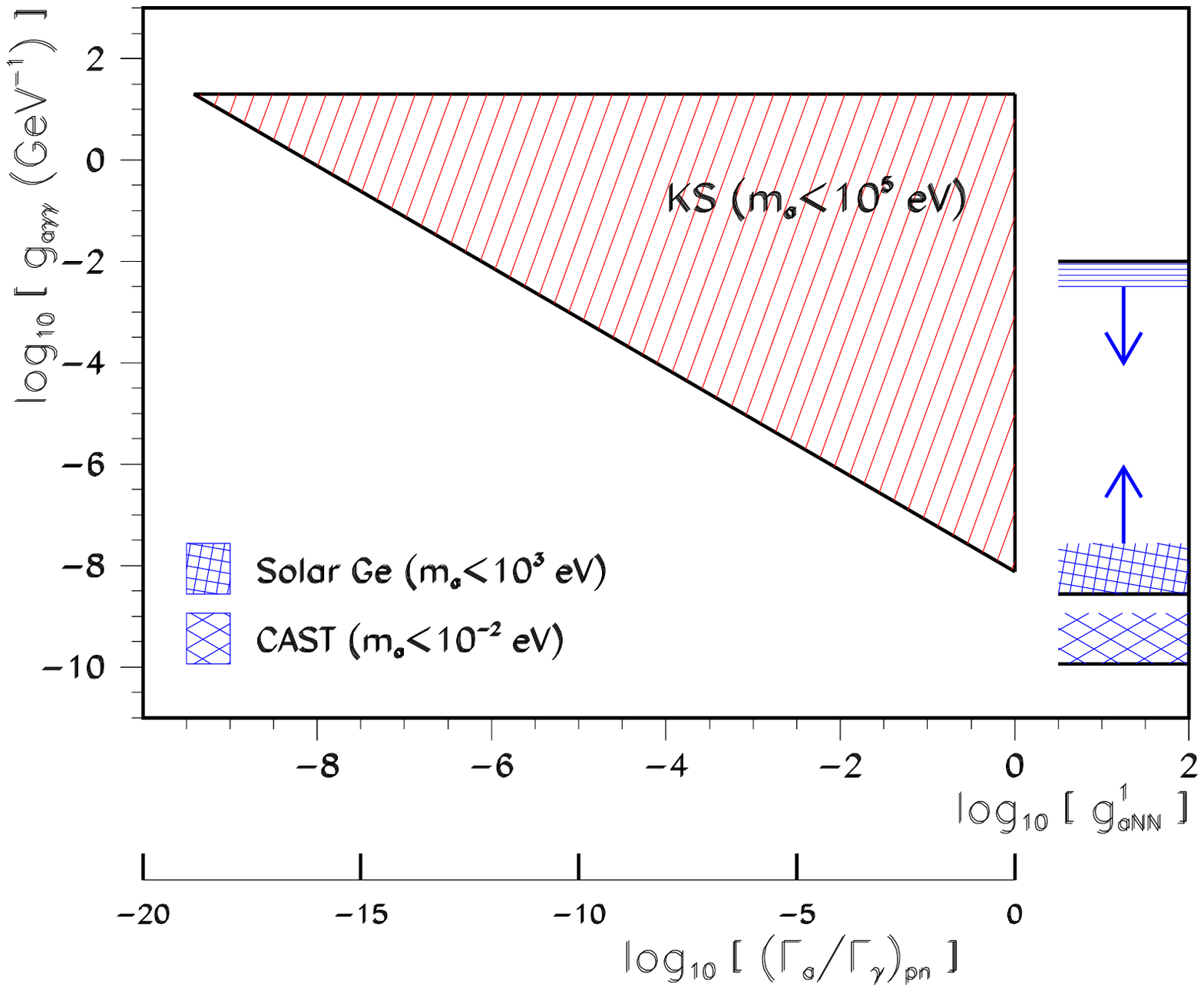}\\
{\bf (b)}\\
\includegraphics[width=8cm]{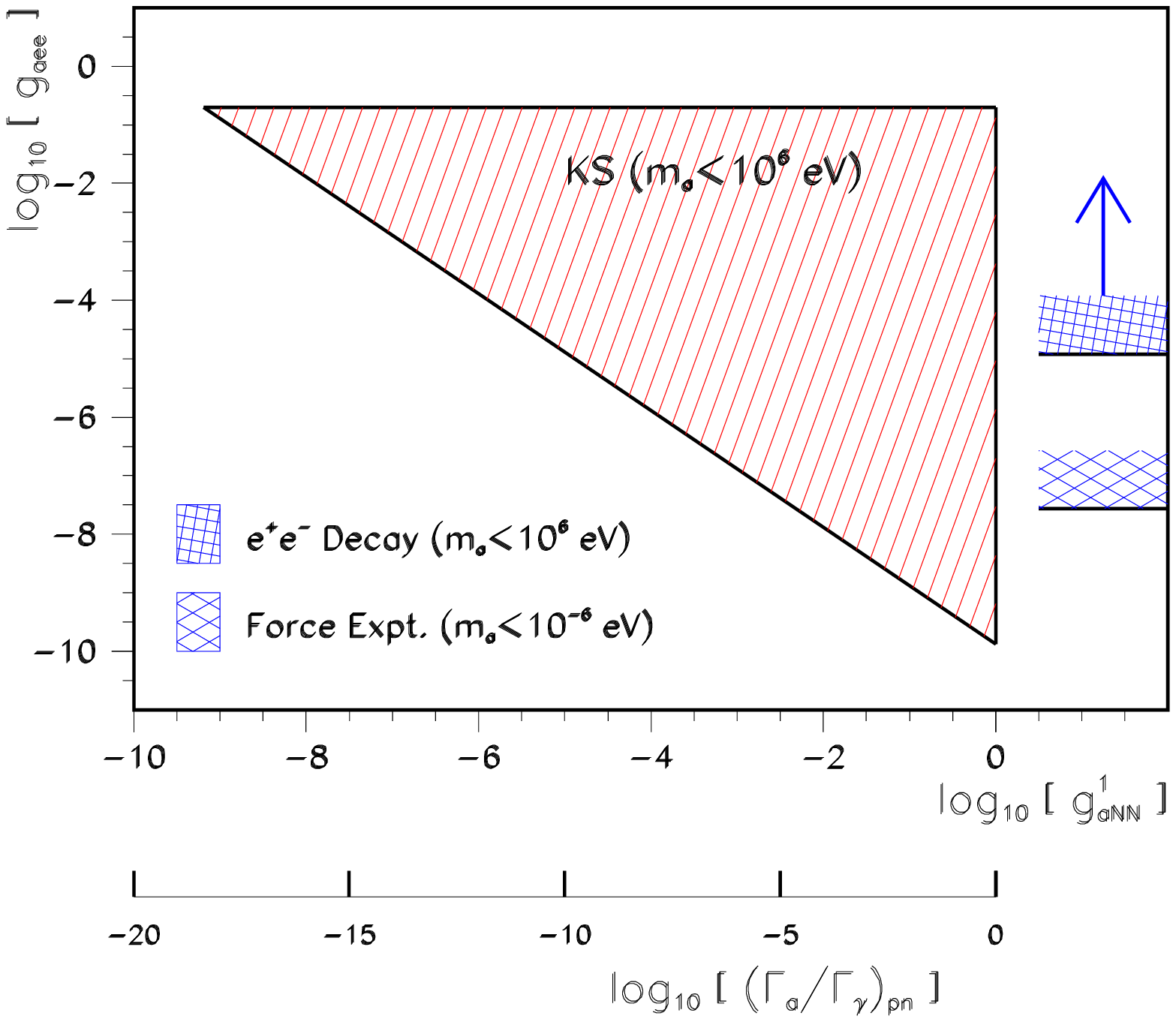}
\caption{
The model-independent exclusion regions of the KS
experiment for (a) $\gagg$ and
(b) $\gaee$  on the  $\braxion _{pn}$
and $\gv1$ axes.
The $\gagg$ and $\gaee$ limits from  
the leading laboratory experiments
are also shown for comparisons.
The ranges of validity in $m_a$ are indicated.
}
\label{g2br}
\end{figure}

The experimental sensitivities are defined by
the signal rates $S _{P/C}$
which are proportional to 
$g^2_{a \gamma \gamma / aee} \braxion$
and alternatively, via Eq.~\ref{eq::pn2dg},
to $g^2_{a \gamma \gamma / aye} \gv1$.
Limit on the  Primakoff conversion rate 
$S _P < S _u ( \pn2dg )$ gives rise to 
the model-independent constraint
\begin{eqnarray}
\multirow{4}*{ \{  }
 & \gagg ^2 \cdot ( \breq ) _{pn}  & ~ < ~~ 
5.9 \times 10^{-17}~ {\rm GeV}^{-2} 
\multirow{4}*{ ~ , } \\
 & \gagg \cdot \gv1 &  ~ < ~~  7.7 \times 10^{-9}~ {\rm GeV ^{-1} }  
\end{eqnarray}
which are applicable within 
the ranges of $m _a \alt 10^5 ~ {\rm eV}$ 
and $\gagg \alt 20 ~ {\rm GeV^{-1}}$, 
bounded by kinematics and axion interactions in flight, respectively.
Similarly,
the limit on the Compton conversion rate
$S _C < S _u ( \pn2dg )$ leads to 
\begin{eqnarray}
\multirow{4}*{ \{  }
& \gaee ^2 \cdot ( \breq ) _{pn}  & ~ < ~ ~  1.7 \times 10^{-20} \\
& \gaee \cdot \gv1  & ~ < ~ ~ 1.3 \times 10^{-10} 
\end{eqnarray}
for $m _a \alt 10^6 ~ {\rm eV}$ and
$\gaee \alt 0.2$.
The loss of sensitivities in $\gagg$ at $m_a \sim 10^6 ~ {\rm eV}$ 
can be explained by
the reduction of the Primakoff cross-section
as depicted in Figure~\ref{csvsma}.
These constraints on $\gagg$ and $\gaee$ as
functions of $\gv1$ and $\braxion _{pn}$
are illustrated respectively 
in Figures.~\ref{g2br}a\&b. 
Limits from the most sensitive laboratory
experiments are
also displayed for comparisons.

The KS limits on $\gagg$
are not as sensitive as those of CAST~\cite{cast} and
the solar-germanium~\cite{solarge} experiments for
$m_a \alt 10^4 ~ {\rm eV}$ at all branching ratios.
On the contrary,
the $\gaee$ sensitivities exceed those
of the positronium decay~\cite{asai} 
and macroscopic force~\cite{force}
experiments.
New regions are probed for 
$m_a \agt 10^{-6} ~ {\rm eV}$
with
$\braxion _{pn} >  10^{-9}$
and  for
$m_a \alt 10^{-6} ~ {\rm eV}$
with
$\braxion _{pn} >  10^{-5}$.

\section{VI. Summary and Prospects}

This article reports the first study
of possible emissions of axions from
power reactors using Primakoff and
Compton conversions as the detection
mechanisms. 
No evidence were observed and
constraints on axion parameters were placed.

The exclusion regions
in the $( m_a , \gagg / \gaee )$
parameter space were identified
within the framework of
the invisible axion models.
The branching ratios for axion emissions
associated with radioactive $\gamma$-decays 
are proportional to $m _a ^2$, such
that the experiment is sensitive mostly at
the large axion mass region.
The KS results define the global exclusion
boundaries for $\gagg$ and $\gaee$ 
and excluded
the KSVZ and DFSZ models for
$10^4 ~{\rm eV} \alt m _a \alt 10^6 ~ {\rm eV}$.

Independent of models,
the KS results are not as sensitive
in constraining $\gagg$ compared
to those from solar axion searches, 
but improve on the limits in $\gaee$ for
branching ratios as small as 
$\braxion _{pn} > 10^{-9}$
for light-mass axions such as those
within the cosmologically-preferred range
of $\rm{10^{-6} ~ eV < m_{a} < 10^{-2} ~ eV}$.

Our studies therefore indicate that this
experimental approach can 
provide competitive sensitivities
compared to the other techniques when
(a) the axion physics is correctly described
by the invisible axion models and
the axion mass is at the keV$-$MeV range, 
or
(b) the axion physics
allows relatively large axion-nucleon couplings
(and consequently the axion emission 
branching ratios) at small axion mass 
not yet covered by the current 
theoretical modeling.

\section{VII. Acknowledgments}

The authors are grateful to
inspiring discussions with
Profs. C.Y.~Chang, H.Y.~Cheng,
L.~Hall, K.B.~Luk and K.W.~Ng.
Suggestions from the referees
are warmly appreciated.
This work is supported by  
contracts 93-2112-M-001-030,
94-2112-M-001-028 and
95-2119-M-001-028
from the National Science Council, Taiwan.



\begin{thebibliography}{99}

\bibitem{pdg}
See
{\it Review of Particle Physics},
W.M.~Yao et al.,
J. Phys. {\bf G 33}, 417-424 (2006),
for details and references.

\bibitem{avignone}
F.T.~Avignone III et al., Phys. Rev. {\bf D 37}, 618 (1988).

\bibitem{pqww}
R.~D.~Pecci and H.~R.~Quinn, Phys. Rev. Lett. {\bf 38}, 1440 (1977);
R.~D.~Pecci and H.~R.~Quinn, Phys. Rev. {\bf D 16}, 1791 (1977).
S.~Weinberg, Phys. Rev. Lett. {\bf 40}, 223 (1978);
F.~Wilczek, Phys. Rev. Lett. {\bf 40}, 279 (1978).

\bibitem{exptreview}
D.~Kinion, I.G.~Irastorza and K.~van~Bibber,
Nucl. Phys. {\bf B} (Proc. Suppl.) {\bf 143}, 417 (2005).

\bibitem{dfsz}
A.~R.~Zhitnitsky, Sov. J. Nucl. Phys. {\bf 31}, 260 (1980);
M.~Dine, W.~Fischler and M.~Srednicki, 
Phys. Lett. {\bf B 104}, 199 (1981).

\bibitem{ksvz}
J.~E.~Kim, Phys. Rev. Lett. {\bf 43}, 103 (1979);
M.~A.~Shifman, A.~I.~Vainshtein and 
V.~I.~Zakharov, Nucl. Phys. {\bf B 166}, 493 (1980).

\bibitem{kaplan}
D.B.~Kaplan, Nucl. Phys. {\bf B 260}, 215 (1985);

\bibitem{srednicki}
M. Srednicki, Nucl. Phys. {\bf B 260}, 689 (1985).

\bibitem{raffelt}
G.G.~Raffelt, ``Stars as Laboratories for Fundamental
Physics'', Chapter 14, U. Chicago Press (1996);
G.G.~Raffelt, hep-ph/0611350 (2006).

\bibitem{raxgg}
J.L. Vuilleumier et al., Phys. Lett. {\bf B 101}, 341 (1981);
A.~Zehnder, K.~Gabathuler, and J.L.~Vuilleumier, 
Phys. Lett. {\bf B 110}, 419 (1982);
V.M.~Datar et al., Phys. Lett. {\bf B 114}, 63 (1982);
G.~D.~Alekseev et al., JETP Lett {\bf 36}, 116 (1982);
J.F.~Cavaignac et al., Phys. Lett. {\bf B 121}, 193 (1983);
V.D.~Ananev et al., Sov. J. Nucl. Phys. {\bf 41}, 585 (1985);
S.~N.~Ketov et al., JETP Lett. {\bf 44}, 146 (1986);
H.R.~Koch and O.W.B.~Schult, Nuovo Cim.\ A {\bf 96}, 182 (1986).

\bibitem{raxee}
M.~Altmann et al., Z. Phys. {\bf C 68}, 221 (1995).

\bibitem{donnelly}
T.W.~Donnelly  et al., Phys. Rev. {\bf D 18}, 1607 (1978).


\bibitem{rnue}
B.~Xin et al., Phys. Rev. {\bf D 72}, 012006 (2005).

\bibitem{fyield}
T.R.~England and B.F.~Rider,
Evaluation and Compilation of Fission Yields,
ENDF-349, LA-UR-94-3106 (1993).

\bibitem{texonomagmom}
H.B.~Li et al., Phys. Rev. Lett. {\bf 90}, 131802 (2003);
H.T.~Wong et al., Phys. Rev. {\bf D 75}, 012001 (2007).

\bibitem{texonodaq}
W.P.~Lai et al.,
Nucl. Instrum. Methods {\bf A 465}, 550 (2001).

\bibitem{solarli7}
M.~Kremar~et~al., Phys. Rev. {\bf D 64}, 115016 (2001);
A.V.~Derbin~et~al., JETP Lett. {\bf 81}, 365 (2005).

\bibitem{moriyama}
S. Moriyama, Phys. Rev. Lett. {\bf 75}, 3222 (1995).

\bibitem{barroso}
A.~Barroso and N.C.~Mukhopadhyay,
Phys. Rev. {\bf C 24}, 2382 (1981).

\bibitem{svalue}
A. Airapetian et al., 
Phys. Rev. {\bf D 75}, 012007 (2007).

\bibitem{haxton}
W.C.~Haxton and K.Y.~Lee, Phys. Rev. Lett. {\bf 66}, 2557 (1991).

\bibitem{hall}
L.J.~Hall and T.~Watari, Phys. Rev. {\bf D 70},  115001 (2004).

\bibitem{beamdump}
A.~Konaka et al., Phys. Rev. Lett. {\bf 57}, 659 (1986);
J. D. Bjorken et al., Phys. Rev. {\bf D 38}, 3375 (1988);
A.~Bross et al., Phys. Rev. Lett. {\bf 67}, 2942 (1991).

\bibitem{solarge}
F.T.~Avignone III~et~al.,~Phys.~Rev.~Lett.~{\bf 81}, 5068~(1998).

\bibitem{cast}
K.~Zioutas et al.,
Phys. Rev. Lett. {\bf 94}, 121301 (2005).

\bibitem{cavity}
R.~Bradley et al., Rev. Mod. Phys. {\bf 75}, 777 (2003).

\bibitem{asai}
S.~Asai et al., Phys. Rev. Lett. {\bf 66}, 2440 (1991).

\bibitem{force}
J.E.~Moody and F.~Wilczek, Phys. Rev. {\bf D 30}, 130 (1984);
V.F.~Bobrakov et al., JETP Lett. {\bf 53}, 294 (1991).

\bibitem{laser}
R. Cameron et al., Phys. Rev. {\bf D 47}, 3707 (1993).

\bibitem{pvlas}
E.~Zavattini et al.,
Phys. Rev. Lett. {\bf 96}, 110406 (2006);
A.~Ringwald, J. Phys. Conf. Series {\bf 39}, 197 (2006).

\end{thebibliography}
\end{document}